\documentclass{sapm}

\usepackage{graphicx}
\usepackage{amsfonts}
\usepackage{amssymb}
\usepackage{amsmath}
\usepackage{amsthm}
\usepackage{tikz}
\theoremstyle{plain}
\newtheorem{prop}[theorem]{Propositon}

\jname{STUDIES IN APPLIED MATHEMATICS}
\jvol{00}
\jyear{2015}
\doi{10.1146/((please add article doi))}

\begin{document}

\title[SUSY AKNS: Darboux-B\"acklund transformations and discrete systems]{A supersymmetric AKNS problem and its Darboux-B\"acklund transformations and discrete systems}

\author[L. L. Xue and Q. P. Liu]{ Lingling Xue and Q. P. Liu\thanks{Address for
correspondence: Department of Mathematics, China University of Mining and Technology,
Beijing 100083, P.~R.~China; e-mail: qpl@cumtb.edu.cn }}
\affil{Department of Mathematics, Ningbo University, Ningbo, P.~R.~China, 315211}
\affil{Department of Mathematics, China University of Mining and Technology,\\
Beijing, P.~R.~China, 100083}

\maketitle

\begin{abstract}
In this paper, we consider a supersymmetric AKNS spectral problem. Two elementary and a binary Darboux transformations are constructed. By means of reductions, Darboux and B\"{a}cklund transformations are given for the supersymmetric modified Korteweg-de Vries, sinh-Gordon and nonlinear Schr\"odinger equations. These Darboux and B\"{a}cklund transformations are adopted for the constructions of integrable discrete super systems, and both semi-discrete and fully discrete systems are presented. Also, the continuum limits of the relevant discrete systems are worked out.
\end{abstract}

\section{Introduction}

It is well-known that integrable discretizations for nonlinear systems begun at the early stage of soliton theory and it is now forty years old. Indeed, at the beginning of the seventies of last century, Ablowitz, Ladik, Hirota and others started their investigations on difference systems. After forty years extensive research, a large amount of works have been accumulated and now there exist various approaches to this problem, here we mention: Lax representations, Hirota's bilinear approach, B\"{a}cklund transformations, geometric method, Hamiltonian method, etc. (see \cite{Suris,Nijihoff,levi-winternitz,bobenko} and the references there).

In the soliton theory, Darboux and B\"{a}cklund transformations have been very effective in the study of nonlinear equations \cite{matveev,dl,Gu,cies}. On the one hand, Darboux and B\"{a}cklund transformations are adequate to construct various solutions for a given nonlinear system such as soliton solutions or rational solutions. On the other hand, they may be adopted to create new integrable systems  of both continuous and discrete types.  What we are interested in is  the discretizations of super or supersymmetric integrable systems. As far as we are aware of, Grahovski and Mikhailov are the first to undertake such a research. In a recent paper \cite{sasha},  they constructed  integrable discretizations for a class of nonlinear Schr\"odinger (NLS) equations on Grassmann algebras. Subsequently, collaborated  with Levi we \cite{xll} succeeded in discretizing the supersymmetric Korteweg-de Vries (KdV) equation  and both semi-discrete and fully discrete supersymmetric KdV equations are given. More recently, we considered a generalized super KdV equation and worked out its different B\"acklund-Darboux transformations and discrete systems \cite{xl} (this partially overlaps with the  work of  Zhou \cite{zhou}). Furthermore, the reduction of one of the B\"acklund-Darboux transformations and the corresponding discrete system were considered for Kupershmidt's super KdV equation.

In this paper, we consider a supersymmetric AKNS problem in the same spirits. It will be shown that much like the classical AKNS scheme, this supersymmetric AKNS problem enables us to embed in  some well-known integrable systems such as the supersymmetric sinh-Gordon, supersymmetric modified Korteweg-de Vries (MKdV) and supersymmetric NLS equations. The main aim of the paper is to study this spectral problem and construct its Darboux transformations. These Darboux transformations will lead naturally to the B\"{a}cklund transformations for the supersymmetric systems we are interested in. Furthermore, they allow us to discretize those systems.

This paper is organized as follows. In Section 2, we recall the linear problem or supersymmetric AKNS problem and list down some interesting supersymmetric systems which may be derived from it. In Sections 3 and 4, two elementary Darboux transformations and a binary Darboux transformation are worked out. It is further shown that the binary Darboux transformation allows us to construct the relevant Darboux and B\"{a}cklund transformations for the well-known supersymmetric integrable systems such as supersymmetric MKdV, sinh-Gordon and NLS equations. In Section 5, we use these transformations to construct discrete integrable super systems and various discrete super systems of both semi-discrete and fully discrete systems are found. In particular, discrete versions of supersymmetric MKdV equation are obtained. In Section 6, different continuum limits of the new obtained discrete systems are performed.  The final section summarizes briefly the results.

\section{Linear problem}
Our starting point is the linear spectral problem which is defined in terms of super matrices, by which we mean matrices with entries involving  both bosonic variables  and fermionic variables. For convenience, we introduce an involution on the algebra of super matrices as follows: given
any matrix $A=(a_{ij})_{i,j\in \mathbb{Z}}$, we define  $A^\dag=(a_{ij}^\dag)_{i,j\in \mathbb{Z}}$
and $a_{ij}^\dag=(-1)^{p(a_{ij})}a_{ij}$ with $p(a_{ij})$ denoting  the parity of $a_{ij}$.

For a super matrix in the standard format \cite{manin}
\[W=\left(
\begin{array}{cccc}
A & B \\
C& D
\end{array}
\right),\]
we also recall the definition of its inverse, which is defined by
\[
W^{-1}=\left(
\begin{array}{cccc}
(A-BD^{-1}C)^{-1} & -A^{-1}B(D-CA^{-1}B)^{-1} \\
-D^{-1}C(A-BD^{-1}C)^{-1}& (D-CA^{-1}B)^{-1}
\end{array}
\right),
\]
and its determinant
\[
\text{Ber}(W)=\text{det}(D-CA^{-1}B)^{-1} \text{det}(A)=\text{det}(A-BD^{-1}C)/\text{det}(D).
\]

In this paper, we consider the following linear problem:
\begin{eqnarray}\label{LAXL_snls}
(\mathcal{D}\Psi)={L}\Psi,\quad
{L}=
\left(
  \begin{array}{cccc}
    0 & \alpha & 0  \\
    \beta & 0 &  \lambda \\
    0& \lambda  & 0
  \end{array}
\right),
\end{eqnarray}
where $\alpha=\alpha(x,\theta,t)$ and $\beta=\beta(x,\theta,t)$ are  fermionic superfields depending on a temporal variable $t$ and super spatial variables $(x,\theta)$, ${\cal D}=\partial_\theta+\theta\partial_x$ is the corresponding super derivative, and $\Psi=(f,g, \sigma)^{\text{T}} $ with $f, g$  being  bosonic components and $\sigma$  the fermionic one.
As will be shown below, different specifications of the field variables will lead to different supersymmetric integrable systems, which include the celebrated supersymmetric sinh-Gordon equation, supersymmetric MKdV equation and supersymmetric NLS equation in particular. Thus, the situation is much like the classical AKNS scheme and we refer the spectral problem as the supersymmetric AKNS spectral problem. It is remarked that this linear problem is different from the fully supersymmetric AKNS spectral problem proposed and studied by Morosi and Pizzocchero \cite{mp}.

We also note that the matrix spectral problem may be rewritten as a scalar form
\[
(\partial_x +\beta\alpha -({\cal D} \beta ){\cal D}^{-1}\alpha)g=\lambda^2 g,
\]
which is gauge-equivalent to the following spectral problem
\[
(\partial_x+\phi {\cal D}^{-1}\psi)g=\lambda^2 g,
\]
so-called sAKNS considered by Aratyn and Das \cite{ad}.

Choose the time dependency of superfield $\Psi$ to be
\begin{eqnarray*}
\Psi_{t}={P}\Psi,
\end{eqnarray*}
or
\begin{eqnarray*}
(\mathcal{D}_t\Psi)={S}\Psi,\quad {\cal D}_t=\partial_{\theta_t}+\theta_t\partial_t,
\end{eqnarray*}
where $\theta_t$ is a fermionic variable such that $(t, \theta_t)$ constitutes the super temporal coordinate.
Then the compatibility condition of the linear problems or the zero curvature representation implies
\begin{equation}\label{zcr1}
(\mathcal{D}_t{L})+(\mathcal{D}{S})+{L}^\dag{S}+{S}^\dag{L}=0
\end{equation}
or
\begin{equation}\label{zcr2}
{L}_t-(\mathcal{D}{P})+{L}{P}-{P}^\dag{L}=0.
\end{equation}
By the various choices of the $P$ or $S$, we are provided with a large class of supersymmetric equations, some of the most important ones are listed below.

\noindent
{\bf Case I:} $\beta=\alpha=\mathcal{D}\Phi$

(a) The supersymmetric sinh-Gordon equation \cite{sg1,sg2,kulish}
\begin{eqnarray}\label{sinhG}
\mathcal{D}\mathcal{D}_t\Phi=2\mathrm{i}\sinh(\Phi).
\end{eqnarray}
In this case, we have
\begin{eqnarray*}
{S}=\frac{2\mathrm{i}}{\lambda}
\left(
  \begin{array}{cccc}
    0 & 0 & \sinh(\Phi)  \\
     0 & 0 & \cosh(\Phi)  \\
      \sinh(\Phi) & -\cosh(\Phi) & 0
  \end{array}
\right).
\end{eqnarray*}

(b) The supersymmetric MKdV equation \cite{mathieu,sy}
\begin{eqnarray}\label{smkdv}
\alpha_t=\alpha_{xxx}-3\alpha({\cal D}\alpha)({\cal D}\alpha_x)-3({\cal D}\alpha)^2\alpha_x,
\end{eqnarray}
now we set
{\small
\[
{P}=
\left(
  \begin{array}{cccc}
    \lambda^2\left((\mathcal{D}\alpha)^2+2\alpha_x\alpha\right) & B & A  \\
      B+2\lambda^2(\mathcal{D}\alpha_x)
    & \lambda^6-\lambda^2(\mathcal{D}\alpha)^2
    &  \lambda\mathcal{D}(\alpha_x\alpha)+\lambda^3\alpha(\mathcal{D}\alpha) \\
       2\lambda^3\alpha_x-A
    &  \lambda\mathcal{D}(\alpha_x\alpha)-\lambda^3\alpha(\mathcal{D}\alpha)
    & \lambda^6+2\lambda^2\alpha_x\alpha
  \end{array}
\right),
\]
}
with
\begin{align*}
A&=\lambda(2\alpha(\mathcal{D}\alpha)^2-\alpha_{xx})
+\lambda^3\alpha_{x}-\lambda^5\alpha,\\
B&=(\mathcal{D}\alpha_{xx})-2(\mathcal{D}\alpha)^3-3\alpha_x\alpha(\mathcal{D}\alpha)
-\lambda^2(\mathcal{D}\alpha_{x})+\lambda^4(\mathcal{D}\alpha),
\end{align*}

\noindent
{\bf Case II:}

(c) The generalized supersymmetric NLS equation
\begin{eqnarray}\label{snls}
\alpha_t=-\alpha_{xx}+2\alpha\mathcal{D}((\mathcal{D}\alpha)\beta),\quad \beta_t=\beta_{xx}-2\beta\mathcal{D}((\mathcal{D}\beta)\alpha).
\end{eqnarray}
To have above equation, we require
{\small
\begin{eqnarray*}
P=
\left(
  \begin{array}{cccc}
    \lambda^2\beta\alpha+\mathcal{D}(\beta(\mathcal{D}\alpha))-\alpha\beta_x
  & \lambda^2(\mathcal{D}\alpha)-(\mathcal{D}\beta_x)
  & \lambda\alpha_x-\lambda^3\alpha  \\
     \lambda^2(\mathcal{D}\beta)+(\mathcal{D}\beta_x)
  & \lambda^4-\mathcal{D}(\beta(\mathcal{D}\alpha))+\alpha\beta_x
  &  \lambda\alpha(\mathcal{D}\beta) \\
       \lambda\beta_x+\lambda^3\beta
  &  -\lambda\beta(\mathcal{D}\alpha)
  & \lambda^4+\lambda^2\beta\alpha
  \end{array}
\right).
\end{eqnarray*} }
Further reduction, namely  $\beta=\kappa\alpha^*$ ($\kappa=\pm1$) and $\hat{t}=\mathrm{i}t$, yields the supersymmetric NLS equation  \cite{rk,bd,bd2}
\begin{eqnarray}\label{snlsreduction}
\mathrm{i}\alpha_{\hat{t}}=-\alpha_{xx}+2\kappa\alpha\mathcal{D}((\mathcal{D}\alpha){\alpha^*}).
\end{eqnarray}

In the next section, we will construct Darboux transformations for the linear problem \eqref{LAXL_snls}.
For convenience, we will assume that both field variables and eigenfunctions are also functions of  integer-valued variables $n_1$, $n_2$,....
 The subscripts $_{[i]}$ used in the following denote the shifts on the discrete variables, for example,
$\alpha_{[\pm1]}=\alpha(..., n_1\pm1,...)$,  $\alpha_{[\pm2]}=\alpha(..., n_2\pm1,...)$.

\section{Elementary Darboux transformations}
Consider a gauge transformation
\begin{eqnarray}\label{dt1}
\Psi_{[1]}\equiv W\Psi
\end{eqnarray}
such that the linear problem is covariant, i.e.
\begin{eqnarray}\label{gauge_1_snls}
(\mathcal{D}{\Psi_{[1]}})={L}_{[1]}\Psi_{[1]}, \quad {L}_{[1]}=
\left(
  \begin{array}{cccc}
    0 & \alpha_{[1]} & 0  \\
    \beta_{[1]} & 0 &  \lambda \\
    0& \lambda  & 0
  \end{array}
\right).
\end{eqnarray}
Here $W$ is a matrix whose entries are differential functions of spatial and temporal coordinates and rational in $\lambda$. Taking account of \eqref{LAXL_snls}, the compatibility of the two linear systems \eqref{dt1} and \eqref{gauge_1_snls} yields
\begin{eqnarray}
(\mathcal{D} W)+ W^\dag {L}-{L}_{[1]} W=0,\label{conditon}
\end{eqnarray}
which may imply a B\"{a}cklund transformation for a properly chosen $W$.

To obtain the meaningful results, we consider the gauge matrix $W$ with entries quadratic in $\lambda$. After certain analysis, it is found that $W$ takes the following form
\begin{eqnarray*}
 W=
\left(
  \begin{array}{cccc}
    \lambda^2m_{11}+p_{11} & p_{12} & \lambda n_{13} \\
     p_{21} & \lambda^2m_{33}+p_{22} & \lambda n_{23} \\
    \lambda n_{31} & \lambda n_{32} & \lambda^2m_{33}+p_{33} \\
  \end{array}
\right),
\end{eqnarray*}
where ($m_{kk},p_{ij}$) and $n_{ij}$ are bosonic and  fermionic functions respectively. Substituting $W$ and $L$ into \eqref{conditon}, we are provided a system of equations which reads as
\begin{align}
 n_{13}&=-\alpha_{[1]}m_{33}+\alpha m_{11},\quad\;
n_{31}=-\beta_{[1]}m_{11}+\beta m_{33},\label{n13n31}\\
 p_{12}&=\alpha_{[1]}n_{23}-(\mathcal{D}n_{13}),\quad\;\;\,
p_{21}=(\mathcal{D}n_{31})-n_{32}\beta,\label{p1212p21}\\
n_{32}&=(\mathcal{D}m_{33})-n_{23},\qquad\quad
p_{22}=p_{33}+(\mathcal{D}n_{32})-n_{31}\alpha,\label{n23n32}
\end{align}
and
\begin{align}
(\mathcal{D}m_{11})&=0,\quad(\mathcal{D}p_{33})=0,\quad \quad m_{33,x}\,=m_{33}(\alpha_{[1]}\beta_{[1]}-\alpha\beta),\label{m11}\\
(\mathcal{D}p_{12})\,&=\alpha_{[1]}p_{22}-\alpha p_{11},\qquad
\;(\mathcal{D}p_{21})=\beta_{[1]}p_{11}-\beta p_{22},\label{dp12dp21}\\
(\mathcal{D}p_{11})\,&=\alpha_{[1]}p_{21}-\beta p_{12},\qquad
\;(\mathcal{D}p_{22})=\beta_{[1]}p_{12}-\alpha p_{21}.\label{dp11dp22}
\end{align}
It is easy to see that the equations (\ref{dp12dp21},\ref{dp11dp22}) result in
\begin{eqnarray}
\mathcal{D}(p_{12}p_{21}-p_{11}p_{22})=0.\label{p12p21}
\end{eqnarray}

The simplest nontrivial case corresponds to $m_{11}=1$, $m_{33}=0$ and $p_{33}=1$, which solve \eqref{m11}.    Then \eqref{n13n31} and the first equation of \eqref{n23n32} lead to
\begin{eqnarray*}
 n_{13}=\alpha,\quad n_{31}=-\beta_{[1]},\quad n_{32}=-n_{23}.
\end{eqnarray*}
Taking \eqref{p1212p21} and the second equation of \eqref{n23n32} into consideration, the second equation of \eqref{dp11dp22} supplies us \[
n_{23,x}=\left(\alpha_{[1]}\beta_{[1]}-\alpha\beta\right) n_{23},
\]
and  to resolve it, we set $n_{23}=0$.
Thus
\[ n_{23}=n_{32}=0,\quad p_{12}=-(\mathcal{D}\alpha),\quad
p_{21}=-(\mathcal{D}\beta_{[1]}),\quad
p_{22}=1-\alpha\beta_{[1]}. \]
Then from \eqref{p12p21}, we have
\[
(\mathcal{D}\alpha)(\mathcal{D}\beta_{[1]})-p_{11}(1-\alpha\beta_{[1]})=p_1^2,
\]
which gives
\[
p_{11}=(1+\alpha\beta_{[1]})\left((\mathcal{D}\alpha)(\mathcal{D}\beta_{[1]})-p_1^2\right),
\]
where $p_1$ is a  constant of integration.
Summarizing above discussion, we obtain a Darboux transformation   of  elementary type, named as elementary  Darboux transformation (eDT). Below we list down two elementary Darboux transformations.

{\bf eDT-I:}
\begin{align}\label{db1_snls}
&\Psi_{[1]} = W_{p_1}(\alpha,\beta) \Psi,\\
&W_{p_1}(\alpha,\beta)=
\left(
  \begin{array}{cccc}
    \lambda^2+\left(1+\alpha\beta_{[1]}\right) \left((\mathcal{D}\alpha)(\mathcal{D}\beta_{[1]})-p_1^2\right)
    & -(\mathcal{D}\alpha) & \lambda \alpha  \\
  -(\mathcal{D}\beta_{[1]}) & 1-\alpha\beta_{[1]} &  0 \\
    -\lambda \beta_{[1]} & 0 & 1
  \end{array}
\right).\nonumber
\end{align}
By the compatibility of the two linear systems \eqref{db1_snls} and \eqref{gauge_1_snls}, namely,
\[
\mathcal{D} (W_{p_1}(\alpha,\beta))+  \left(W_{p_1}(\alpha,\beta)\right)^\dag {L}-{L}_{[1]} W_{p_1}(\alpha,\beta)=0,
\]
we arrive at the following  B\"{a}cklund transformation
\begin{subequations}\label{snls_case1bt}
\begin{eqnarray}
&&\alpha_{[1]}\;\;=
-\alpha_x(1+\alpha\beta_{[1]})+\alpha\left((\mathcal{D}\alpha)(\mathcal{D}\beta_{[1]})-p_1^2\right),\\
&&\beta_{[1],x}=\beta(1-\alpha\beta_{[1]})-\beta_{[1]}\left((\mathcal{D}\alpha)(\mathcal{D}\beta_{[1]})-p_1^2\right).
\end{eqnarray}
\end{subequations}

It is well known that the inverse of a Darboux matrix is a Darboux matrix as well. Thus by considering $(W_{p_1}(\alpha,\beta))^{-1}$ and  replacing $p_1$, $\alpha$ and $\beta_{[1]}$  by $p_2$, $\alpha_{[2]}$ and $\beta$  respectively, we have a second elementary Darboux transformation.

{\bf eDT-II:}
{\small
\begin{align}\label{db2_snls}
&\Psi_{[2]}= W_{p_2}(\alpha,\beta)\Psi,\\
& W_{p_2}(\alpha,\beta)=
\left(
  \begin{array}{cccc}
    1-\alpha_{[2]}\beta
    & (\mathcal{D}\alpha_{[2]}) & -\lambda \alpha_{[2]}  \\
  (\mathcal{D}\beta) & (1+\alpha_{[2]}\beta )\left(\lambda^2-p_2^2+(\mathcal{D}\alpha_{[2]})(\mathcal{D}\beta)\right)
  &  -\lambda \alpha_{[2]}(\mathcal{D}\beta)\\
    \lambda \beta & \lambda \beta(\mathcal{D}\alpha_{[2]}) & \lambda^2-p_2^2+\lambda^2\alpha_{[2]}\beta
  \end{array}
\right).\nonumber
\end{align}
}
Similarly, the corresponding compatibility condition yields
\[
\mathcal{D} (W_{p_2}(\alpha,\beta))+ \left(W_{p_2}(\alpha,\beta)\right)^\dag {L}-{L}_{[2]} W_{p_2}(\alpha,\beta)=0
\]
which leads to a B\"{a}cklund transformation
\begin{subequations}\label{snls_case2bt}
\begin{align}
\alpha_{[2],x}&=\alpha(\alpha_{[2]}\beta-1)+\alpha_{[2]}\left((\mathcal{D}\alpha_{[2]})(\mathcal{D}\beta)-p_2^2\right),\\
\beta_{[2]}\;&= \beta_x(1+\alpha_{[2]}\beta)+\beta \left((\mathcal{D}\alpha_{[2]})(\mathcal{D}\beta)-p_2^2 \right).
\end{align}
\end{subequations}


As mentioned already, the Darboux matrix $W_{p_2}(\alpha,\beta)$ is essentially the inverse of the Darboux matrix $W_{p_1}(\alpha,\beta)$, thus strictly speaking the eDT-II can not be considered as a new Darboux transformation. We list it down here for the comparison with the existing literatures and convenience to the subsequent discussions. As a matter of fact, the B\"acklund transformations of elementary type were introduced by Konopelchenko for the nonlinear equations solvable by the classical AKNS spectral problem \cite{konopel} (see also \cite{cd,kr}). The Darboux transformations eDT-I and eDT-II are the supersymmetric extensions of the dressing (gauge) transformations of the classical AKNS spectral problem (see (3.25) and (3.28) of \cite{kr}). Also, by taking bosonic limits,  \eqref{snls_case1bt} and \eqref{snls_case2bt} reduce to those presented in \cite{konopel}  (cf. $B^{(1)}_{\lambda}$, $B^{(2)}_{\mu}$ of \cite{konopel} or (2.2), (2.3) of \cite{cd}). Thus, our elementary B\"{a}cklund transformations \eqref{snls_case1bt} and \eqref{snls_case2bt} are natural supersymmetric generalizations of the early results.

The notations $W_{p_i}(\alpha,\beta)$ we adopted above for the Darboux matrices are precise and self-explanatory, however for clarity we may instead use a short-hand notation $W_i \equiv W_{p_i}(\alpha,\beta)$ in sequel. In case where confusion will arise, we will restore $W_{p_i}(\alpha,\beta)$.

So far we have constructed two elementary Darboux matrices $W_1$ and $W_2$ which are expressed by $\alpha$, $\beta_{[1]}$ and $\alpha_{[2]}$, $\beta$, respectively, these Darboux matrices may be made more explicit by means of the eigenfunctions of the linear problem and its adjoint. Therefore, we now consider the adjoint linear problem and construct its elementary Darboux transformations. The adjoint spectral problem of \eqref{LAXL_snls} is given by
\begin{eqnarray}\label{LAXL2_snls}
-\mathcal{D}(\varphi^\dag)=\varphi{L}.
\end{eqnarray}
Consider the following Darboux transformation
 \[\varphi_{[1]}=\varphi T\]
such that
\begin{eqnarray}\label{conditon2}
\mathcal{D}(T^\dag)=L T^\dag-T L_{[1]} ,
\end{eqnarray}
which gives a B\"acklund transformation for a proper $T$.
One can easily check that $T=(W^{-1})^\dag$ satisfies \eqref{conditon2} provided $W$ is a Darboux matrix for \eqref{LAXL_snls},
thus from $W_1$ and $W_2$, we can obtain the corresponding Darboux transformations of the adjoint spectral problem. Indeed, defining  $T_1\equiv T_{p_1}(\alpha,\beta)\equiv (\lambda^2-p_1^2)\left({W^{-1}_{p_1}(\alpha,\beta)}\right)^\dag$, we have explicitly
{\small
\begin{align}\label{T1}
&\varphi_{[1]}= \varphi T_1,\\
&
T_{p_1}(\alpha,\beta)=
\left(
  \begin{array}{cccc}
    1-\alpha\beta_{[1]}
    & (\mathcal{D}\alpha) & \lambda \alpha  \\
  (\mathcal{D}\beta_{[1]}) & (1+\alpha\beta_{[1]} )\left(\lambda^2-p_1^2+(\mathcal{D}\alpha)(\mathcal{D}\beta_{[1]})\right)
  &  \lambda \alpha(\mathcal{D}\beta_{[1]})\\
    -\lambda \beta_{[1]} & -\lambda \beta_{[1]}(\mathcal{D}\alpha) & \lambda^2(1+\alpha\beta_{[1]} )-p_1^2
  \end{array}
\right)\nonumber
\end{align}
}
and the associated  B\"acklund transformation  \eqref{snls_case1bt}. Similarly, letting $T_2\equiv T_{p_2}(\alpha,\beta)\equiv (\lambda^2-p_2^2)\left({W^{-1}_{p_2}(\alpha,\beta)}\right)^\dag$, we find
{\small
 \begin{align}\label{T2}
&\varphi_{[2]}= \varphi T_2,\\
& T_{p_2}(\alpha,\beta)=
\left(
  \begin{array}{cccc}
    \lambda^2+\left(1+\alpha_{[2]}\beta\right) \left((\mathcal{D}\alpha_{[2]})(\mathcal{D}\beta)-p_2^2\right)
    & -(\mathcal{D}\alpha_{[2]}) & -\lambda \alpha_{[2]}  \\
  -(\mathcal{D}\beta) & 1-\alpha_{[2]}\beta &  0 \\
    \lambda \beta & 0 & 1
  \end{array}
\right)\nonumber
\end{align}
}
and  the related B\"acklund transformation \eqref{snls_case2bt}.

Next we intend to work out the explicit representations for our Darboux matrices. By calculating the determinants of $W_1$ and $T_2$, we obtain
\[
\text{Ber}\left(W_1\right)=\lambda^2-p_1^2,\;\; \text{Ber}\left(T_2\right)=\lambda^2-p_2^2.
\]
Now suppose that $\Psi_1=(f_1, g_1, \sigma_1)^{\texttt{T}}$ is a solution of  \eqref{LAXL_snls} at $\lambda=p_1$ and $\varphi_2=(f_2, g_2, \sigma_2)$ solves  \eqref{LAXL2_snls} for $\lambda=p_2$.
Here $f_i$ and $g_i$ are bosonic, $\sigma_i$ are fermionic.
 For convenience, we denote
  $y_i=\frac{g_i}{f_i},\;
\Omega_i=\frac{\sigma_i}{f_i},\; i=1,2$.

As a result, $\beta_{[1]}$ and $\alpha_{[2]}$ are  functions to be expressed by $\Omega_1$ and $\Omega_2$, respectively. Indeed, it follows from the condition $(W_1\Psi)|_{\lambda=p_1}=0$  that
\begin{eqnarray}\label{beta1}
\beta_{[1]} \equiv \frac{\Omega_1}{p_1},\quad (\mathcal{D}\beta_{[1]}) =y_1\left(1-\frac{\alpha\Omega_1}{p_1}\right).
\end{eqnarray}
We can easily check that the second equation of \eqref{beta1} is the consequence of the first one, therefore the first one defines $\beta_{[1]}$. Meanwhile (\ref{snls_case1bt}a) or
\begin{eqnarray}
\alpha_{[1]}\equiv
-\alpha_x\left(1+\frac{\alpha\Omega_1}{p_1}\right)+\alpha\left[y_1(\mathcal{D}\alpha) -p_1^2\right]
\end{eqnarray}
specifies  $\alpha_{[1]}$.

In a similar manner, it follows from the condition $\left(\varphi T_2\right)|_{\lambda=p_2}=0$ that
\begin{eqnarray}\label{alpha2}
\alpha_{[2]} \equiv \frac{\Omega_2}{p_2},\quad (\mathcal{D}\alpha_{[2]}) =y_2\left(1+\frac{\beta\Omega_2}{p_2}\right).
\end{eqnarray}
Also the first equation of \eqref{alpha2} defines $\alpha_{[2]}$ completely and (\ref{snls_case2bt}b), that is,
\begin{eqnarray}\label{beta2}
\beta_{[2]}\equiv \beta_x\left(1+\frac{\Omega_2\beta}{p_2}\right)+\beta \left[y_2(\mathcal{D}\beta)-p_2^2 \right]
\end{eqnarray}
defines $\beta_{[2]}$.

With above results, we arrive at the explicit Darboux matrices which are listed below
{\small
\begin{align}\label{exDT1}
&W_{p_1}(\alpha,\beta)=\begin{pmatrix}\lambda^2-p_1^2+y_1(\mathcal{D}\alpha)-p_1\alpha\Omega_1&-(\mathcal{D}\alpha)&\lambda\alpha\\
-y_1\left(1-\frac{\alpha\Omega_1}{p_1}\right)&1-\frac{\alpha\Omega_1}{p_1}&0\\
-\lambda\frac{\Omega_1}{p_1}&0&1\end{pmatrix},\\
&\label{exDT2}
W_{p_2}(\alpha,\beta)=\begin{pmatrix}1-\frac{\Omega_2\beta}{p_2}&y_2\left(1+\frac{\beta\Omega_2}{p_2}\right)&-\lambda\frac{\Omega_2}{p_2}\\[5pt]
(\mathcal{D}\beta) &(\lambda^2-p_2^2)\left(1+\frac{\Omega_2\beta}{p_2}\right)+y_2(\mathcal{D}\beta) &-\lambda\frac{(\mathcal{D}\beta)\Omega_2}{p_2}\\[5pt]
\lambda\beta&\lambda\beta y_2&\lambda^2\left(1+\frac{\Omega_2\beta}{p_2}\right)-p_2^2\end{pmatrix},
\\
& \label{exDT4}
T_{p_i}(\alpha,\beta)= (\lambda^2-p_i^2)\left({W^{-1}_{p_i}(\alpha,\beta)}\right)^\dag, \; i=1, 2.
\end{align}
}
Our findings on the elementary Darboux transformations are summarized in the following proposition.
\begin{prop}
Given a solution $\Psi_1=(f_1, g_1, \sigma_1)^{\texttt{T}}$ of  \eqref{LAXL_snls} at $\lambda=p_1$
and a solution $\varphi_2=(f_2, g_2, \sigma_2)$ of  \eqref{LAXL2_snls} at $\lambda=p_2$, then
linear problem \eqref{LAXL_snls} and its adjoint linear problem \eqref{LAXL2_snls} admit two elementary Darboux transformations, respectively, i.e.
\[
\Psi_{[i]} = W_{p_i}(\alpha,\beta)\Psi, \quad \varphi_{[i]} = \varphi T_{p_i}{(\alpha,\beta)},\quad i=1,\;2,
\]
where $W_{p_i}(\alpha,\beta)$ and $T_{p_i}(\alpha,\beta) $ are defined by (\ref{exDT1}-\ref{exDT4})
with $\alpha_{[i]}$ and  $\beta_{[i]}$ given by (\ref{beta1}-\ref{beta2}).
\end{prop}


\section{ Binary Darboux transformation}


The elementary Darboux transformations constructed in last section are relevant and useful for the study of the general nonlinear systems associated with the linear problem, but it is not known how to restrict them to the interesting reductions such as supersymmetric NLS or MKdV equations.
In this section, we first demonstrate the permutation property of our elementary Darboux transformations, then we construct a binary Darboux transformation, which may be easily implemented for the specific reductions.

The composition of the  elementary Darboux transformations may be done in two different ways as depicted   in Figure 1 below.
\begin{figure}[htb]
\centering\begin{tikzpicture}
\draw [->](0.5,0.3)--(1.7,0.8);
\draw [->](0.5,-0.3)--(1.7,-0.8);
\draw [->](3.5,-0.8)--(4.7,-0.4) node [right] {$(\Psi_{[21]},\varphi_{[21]})$};
\draw [->](3.5,0.8)--(4.7,0.4) node [right] {$(\Psi_{[12]},\varphi_{[12]})$};
\node at ( 0,0) {$(\Psi,\varphi)$};
\node at ( 2.6,1)  {$(\Psi_{[1]},\varphi_{[1]})$};
\node at ( 2.6,-1)  {$(\Psi_{[2]},\varphi_{[2]})$};
\node at ( 1.1,0.58) [above] {$p_1$};
\node at ( 1.15,-0.55) [below] {$p_2$};
\node at ( 4.1,0.58) [above] {$p_2$};
\node at ( 4.1,-0.55) [below] {$p_1$};
\end{tikzpicture}
  \caption{Iteration of Elementary DTs.}
\end{figure}
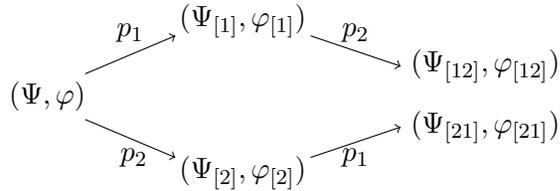
\\
Therefore, on the one hand we have
{\small
\begin{align*}
&\Psi_{[21]}= W_{p_1}(\alpha_{[2]},\beta_{[2]})W_{p_2}(\alpha,\beta)\Psi,\\
&
W_{p_1}(\alpha_{[2]},\beta_{[2]}) =
\left(
  \begin{array}{cccc}
    \lambda^2+\left(1+\alpha_{[2]}\beta_{[21]}\right) \left((\mathcal{D}\alpha_{[2]})(\mathcal{D}\beta_{[21]})-p_1^2\right)
    & -(\mathcal{D}\alpha_{[2]}) & \lambda \alpha_{[2]}  \\
  -(\mathcal{D}\beta_{[21]}) & 1-\alpha_{[2]}\beta_{[21]} &  0 \\
    -\lambda \beta_{[21]} & 0 & 1
  \end{array}
\right),
\end{align*}
}
with
\begin{eqnarray}\label{alpha21}
\alpha_{[21]}=
-\alpha_{[2],x}(1+\alpha_{[2]}\beta_{[21]})+\alpha_{[2]}\left((\mathcal{D}\alpha_{[2]})(\mathcal{D}\beta_{[21]})-p_1^2\right).
\end{eqnarray}
To find an expression for $\beta_{[21]}$, we consider the equation $\left(W_{p_1}(\alpha_{[2]},\beta_{[2]})W_{p_2}(\alpha,\beta)\Psi
\right)|_{\lambda=p_1}=0$  which yields
{\small
\begin{align}\label{beta21}
\beta_{[21]}\;\;&\equiv  \beta+\frac{( p_1^2- p_2^2)\Omega_1\left({p_2}-\beta\Omega_2\right)}{{p_1}{p_2}(1+y_1y_2)},\\
(\mathcal{D}\beta_{[21]})&=(\mathcal{D}\beta)+\frac{p_1^2-p_2^2}{p_1p_2(1+y_1y_2)}\left[(p_2-\beta\Omega_2)p_1y_1
+\left((\mathcal{D}\beta)-\frac{p_2^2y_1}{1+y_1y_2}\right)\Omega_1\Omega_2\right].\nonumber
\end{align}
}
A direct calculation shows that above two equations are consistent,  therefore \eqref{beta21} defines $\beta_{[21]}$. And
from \eqref{alpha21},  we find
\[
\alpha_{[21]}\equiv \alpha-\frac{( p_1^2- p_2^2)\Omega_2\left({p_1}+\alpha\Omega_1\right)}{{p_1}{p_2}(1+y_1y_2)}.
\]

On the other hand, we also have
{\small
\begin{align*}
&\Psi_{[12]}= W_{p_2}(\alpha_{[1]},\beta_{[1]})W_{p_1}(\alpha,\beta)\Psi,\\
&W_{p_2}(\alpha_{[1]},\beta_{[1]}) =\\
&\left(
  \begin{array}{cccc}
    1-\alpha_{[12]}\beta_{[1]}
    & (\mathcal{D}\alpha_{[12]}) & -\lambda \alpha_{[12]}  \\
  (\mathcal{D}\beta_{[1]}) & (1+\alpha_{[12]}\beta_{[1]} )\left(\lambda^2-p_2^2+(\mathcal{D}\alpha_{[12]})(\mathcal{D}\beta_{[1]})\right)
  &  -\lambda \alpha_{[12]}(\mathcal{D}\beta_{[1]})\\
    \lambda \beta_{[1]} & \lambda \beta_{[1]}(\mathcal{D}\alpha_{[12]}) & \lambda^2(1+\alpha_{[12]}\beta_{[1]} )-p_2^2
  \end{array}
\right)
\end{align*}
}
and
{\small
 \begin{align*}
&\varphi_{[12]}= \varphi T_{p_1}(\alpha,\beta)T_{p_2}(\alpha_{[1]},\beta_{[1]}),\\
&T_{p_2}(\alpha_{[1]},\beta_{[1]})=\nonumber\\
&\left(
  \begin{array}{cccc}
    \lambda^2+\left(1+\alpha_{[12]}\beta_{[1]}\right) \left((\mathcal{D}\alpha_{[12]})(\mathcal{D}\beta_{[1]})-p_2^2\right)
    & -(\mathcal{D}\alpha_{[12]}) & -\lambda \alpha_{[12]}  \\
  -(\mathcal{D}\beta_{[1]}) & 1-\alpha_{[12]}\beta_{[1]} &  0 \\
    \lambda \beta_{[1]} & 0 & 1
  \end{array}
\right)\nonumber
\end{align*}
}
with
\begin{eqnarray}
\label{beta12}
\beta_{[12]}\equiv \beta_{[1],x}(1+\alpha_{[12]}\beta_{[1]})+\beta_{[1]} \left((\mathcal{D}\alpha_{[12]})(\mathcal{D}\beta_{[1]})-p_2^2 \right).
\end{eqnarray}
The equation $\left(\varphi T_{p_1}(\alpha,\beta)T_{p_2}(\alpha_{[1]},\beta_{[1]})\right)|_{\lambda=p_2}=0$  leads to
{\small\begin{equation*}
\begin{split}
\alpha_{[12]}\;\;&\equiv \alpha-\frac{(p_1^2-p_2^2)\Omega_2(p_1+\alpha\Omega_1}{p_1p_2(1+y_1y_2)},\label{alpha12}\\
(\mathcal{D}\alpha_{[12]})&= (\mathcal{D}\alpha)-\frac{p_1^2-p_2^2}{p_1p_2(1+y_1y_2)}\left[(p_1+\alpha\Omega_1)p_2y_2
+\left((\mathcal{D}\alpha)-\frac{p_1^2y_2}{1+y_1y_2}\right)\Omega_1\Omega_2\right].
\end{split}
\end{equation*}}
%
Once again, it is straightforward to check the consistency of last two equations, thus the first defines $\alpha_{[12]}$.
And \eqref{beta12},  i.e.
\[
\beta_{[12]}\equiv \beta+\frac{( p_1^2- p_2^2)\Omega_1\left({p_2}-\beta\Omega_2\right)}{{p_1}{p_2}(1+y_1y_2)},
\]
defines $\beta_{[12]}$.

Based on above discussions, we achieve
\[\alpha_{[3]}\equiv\alpha_{[12]}=\alpha_{[21]},\quad\beta_{[3]}\equiv\beta_{[12]}=\beta_{[21]},\]
from which we can show  that
\begin{align*}
\Psi_{[3]}&\equiv\Psi_{[12]}=\Psi_{[21]}, \\
  W_3&\equiv W_{p_1}(\alpha_{[2]},\beta_{[2]})W_{p_2}(\alpha,\beta)=W_{p_2}(\alpha_{[1]},\beta_{[1]})W_{p_1}(\alpha,\beta).
  \end{align*}
Hence these two elementary Darboux transformations do commute and the composition of them generates a compound or  binary transformation which is summarized in the following proposition.
\begin{prop}
Linear problem \eqref{LAXL_snls}     admits a binary Darboux transformation,  that is,
\begin{align*}
\Psi_{[3]}&\equiv W_3\Psi,\quad
 W_3=
\left(
  \begin{array}{cccc}
    \lambda^2+p_{11} & p_{12} & \lambda n_{13} \\
     p_{21} & \lambda^2(1+\Delta)+p_{22} & \lambda n_{23} \\
    \lambda n_{31} & \lambda n_{32} & \lambda^2(1+\Delta)- p_2^2 \\
  \end{array}
\right),\\
\alpha_{[3]}&\equiv \alpha-\frac{2k_1\Omega_2\left({p_1}+\alpha\Omega_1\right)}{{p_1}{p_2}(1+y_1y_2)},\qquad
\beta_{[3]}\equiv \beta+\frac{2k_1\Omega_1\left({p_2}-\beta\Omega_2\right)}{{p_1}{p_2}(1+y_1y_2)},
\end{align*}
where
\begin{align*}
k_1\;&=\frac{1}{2}( p_1^2- p_2^2),\qquad\qquad\quad\qquad\quad\;\;
\Delta\;\,=\frac{2k_1\Omega_1\Omega_2}{{p_1}{p_2}(1+y_1y_2)},\\
n_{23}&=y_1 n_{13},\qquad\qquad\qquad\qquad\quad\;\;\quad\,
n_{32}=y_2 n_{31},\\
n_{13}&=\frac{2k_1\Omega_2}{{p_2}(1+y_1y_2)},\qquad\qquad\qquad\quad\;\;
n_{31}=\frac{-2k_1\Omega_1}{{p_1}(1+y_1y_2)},\\
p_{12}&=\frac{2k_1 y_2}{-1-y_1y_2}
       +\frac{2{p_1}k_1y_2\Omega_1\Omega_2}{{p_2}(1+y_1y_2)^2},\quad\;\;
p_{21}=\frac{2k_1 y_1}{-1-y_1y_2}
       +\frac{2 {p_2}k_1y_1\Omega_1\Omega_2}{{p_1}(1+y_1y_2)^2},\\
p_{11}&=\frac{ p_1^2+ p_2^2y_1y_2}{-1-y_1y_2}
       +\frac{2{p_1}k_1\Omega_1\Omega_2}{{p_2}(1+y_1y_2)^2},\quad
p_{22}=\frac{ p_2^2+ p_1^2y_1y_2}{-1-y_1y_2}
       -\frac{2{p_2}k_1\Omega_1\Omega_2}{{p_1}(1+y_1y_2)^2}.
\end{align*}
\end{prop}
In order to derive the corresponding B\"acklund transformation,
we eliminate $y_i$ and $\Omega_i$ by the transformations
{\small
\begin{eqnarray*}
&&\Omega_1 =\frac{p_1}{2h(h+k_1)}(2h+(\alpha_{[3]}-\alpha)\beta )(\beta_{[3]}-\beta),\\
&&\Omega_2 =\frac{-p_2}{2h(h+k_1)}(2h+(\beta_{[3]}-\beta)\alpha)(\alpha_{[3]}-\alpha),\\
&&y_1=\frac{\mathcal{D}(\beta_{[3]}-\beta)}{2h(h+{k_1})}\left[2h+(\alpha_{[3]}-\alpha)\beta+\frac{k_1-h}{h+k_1}(\beta_{[3]}-\beta)\alpha\right.\\
&&\qquad\left.+\frac{k_1^2+2hk_1-h^2}{2h^2(h+k_1)}\alpha\beta\alpha_{[3]}\beta_{[3]}\right]
-\frac{(\alpha_{[3]}-\alpha)(\beta_{[3]}-\beta)}{2h(h+k_1)^3}\\
&&\qquad\times\left[(\mathcal{D}\beta)(h+k_1)^2
+\mathcal{D}(\beta_{[3]}-\beta)(h^2-hp_2^2+k_1k_2+\mathcal{D}(\beta_{[3]}-\beta)(\mathcal{D}\alpha))\right],\\
&&y_2=\frac{\mathcal{D}(\alpha-\alpha_{[3]})}{2h(h+{k_1})}\left[2h+(\beta_{[3]}-\beta)\alpha+\frac{k_1-h}{h+k_1}(\alpha_{[3]}-\alpha)\beta\right.\\
&&\qquad\left.+\frac{k_1^2+2hk_1-h^2}{2h^2(h+k_1)}\alpha\beta\alpha_{[3]}\beta_{[3]}\right]
-\frac{(\alpha_{[3]}-\alpha)(\beta_{[3]}-\beta)}{2h(h+k_1)^3}\\
&&\qquad\times\left[(\mathcal{D}\alpha)(h+k_1)^2
+\mathcal{D}(\alpha_{[3]}-\alpha)(h^2+hp_1^2-k_1k_2+\mathcal{D}(\alpha_{[3]}-\alpha)(\mathcal{D}\beta))\right],
\end{eqnarray*}
}
where
\begin{eqnarray*}
k_2=-\frac{1}{2}\left(p_1^2+ p_2^2\right),\quad
h\equiv\sqrt{k_1^2+\mathcal{D}(\alpha_{[3]}-\alpha)\mathcal{D}(\beta_{[3]}-\beta)}.
 \end{eqnarray*}
In this way,  one obtains
\begin{eqnarray*}
&&\Delta\;=\frac{1}{{k_1}+h}(\alpha_{[3]}-\alpha)(\beta_{[3]}-\beta),\\
&&n_{13}=\alpha-\alpha_{[3]}-\alpha \Delta,\qquad\quad\; n_{31}=\beta-\beta_{[3]}+\beta \Delta,\\
&&n_{23}=-\frac{\mathcal{D}(\beta_{[3]}-\beta)}{2 h(h+{k_1})}\left[2h-\alpha(\beta_{[3]}-\beta)\right]\left(\alpha_{[3]}-\alpha\right),\\
&&n_{32}=\frac{\mathcal{D}(\alpha_{[3]}-\alpha)}{2 h(h+{k_1})}\left[2h-\beta(\alpha_{[3]}-\alpha)\right]\left(\beta_{[3]}-\beta\right),\\
&&p_{12}=\mathcal{D}(\alpha_{[3]}-\alpha)
-\frac{\beta_{[3]}-\beta}{h+{k_1}}\mathcal{D}(\alpha\alpha_{[3]})
+\frac{\Delta}{2 h}\mathcal{D}(\alpha_{[3]}-\alpha)\alpha\beta,\\
&&p_{21}=\mathcal{D}(\beta-\beta_{[3]})
+\frac{\alpha_{[3]}-\alpha}{h+{k_1}}\mathcal{D}(\beta\beta_{[3]})
+\frac{\Delta}{2 h}\mathcal{D}(\beta_{[3]}-\beta)\beta\alpha,\\
&&p_{11}=k_2-h-\Delta_{11}-\Delta_{41},\qquad
p_{22}=k_2+h+\Delta_{22}+\Delta_{41},
\end{eqnarray*}
where
\begin{eqnarray*}
&&\Delta_{41}=\frac{1}{4 h^3}\left(h-{k_1}\right)\left(k_1^2+2 {k_1} h+3h^2\right)\alpha\beta\Delta,\\
&&\Delta_{22}=\Delta_{11}+\left(\alpha_{[3]}-\alpha\right)\left( \beta_{[3]}-\beta\right),\\
&&\Delta_{11}=
\frac{\Delta}{2h} \left[ (\mathcal{D}\beta_{[3]})(\mathcal{D}\alpha)-(\mathcal{D}\alpha_{[3]})(\mathcal{D}\beta)\right]\\
&&\qquad\quad+\frac{h-{k_1}}{2h}\left(\alpha_{[3]}\beta+\beta_{[3]}\alpha\right)
-\frac{h-k_2}{2h}\left(\alpha_{[3]}
-\alpha\right)\left( \beta_{[3]}-\beta\right).
\end{eqnarray*}
The compatibility condition is
\begin{eqnarray*} 
(\mathcal{D} W_3)+ W_3^\dag {L}-{L}_{[3]} W_3=0,
\end{eqnarray*}
which leads to a B\"{a}cklund transformation
\begin{eqnarray*}
&&\alpha_{[3],x}=\alpha_x +
 (k_1+k_2)\alpha_{[3]}
+\alpha \left(h-k_2\right)\nonumber\\
&&\qquad\quad\;+\frac{(\mathcal{D}\alpha_{[3]})}{{k_1}+h}(\alpha_{[3]}-\alpha )\mathcal{D}(\beta_{[3]}-\beta)\nonumber\\
&&\qquad\quad\;-(\alpha_{[3]}-\alpha)(\beta_{[3]}-\beta)\left[ \frac{\alpha_{x}}{{k_1}+h}
+\frac{h-k_2}{2h}\alpha \right]\nonumber\\
&&\qquad\quad\;+\frac{\alpha_{[3]}\alpha\beta}{2h({k_1}+h)}
\left[
2(\mathcal{D}\alpha)(\mathcal{D}\beta_{[3]})-(\mathcal{D}\beta)\mathcal{D}(\alpha_{[3]}+\alpha))\right]\nonumber\\
&&\qquad\quad\;+\frac{\alpha_{[3]}\alpha\beta_{[3]}}{2h({k_1}+h)}
   \left[2(\mathcal{D}\alpha_{[3]})(\mathcal{D}\beta )-(\mathcal{D}\beta_{[3]})\mathcal{D}(\alpha_{[3]}+\alpha)
          \right],
\end{eqnarray*}

\begin{eqnarray*}
&&\beta_{[3],x}=\beta_x+  (k_1-k_2)\beta_{[3]}+\beta\left(h+k_2\right)\nonumber\\
&&\qquad\quad\;+\frac{(\mathcal{D}\beta_{[3]})}{{k_1}+h}(\beta_{[3]}-\beta )\mathcal{D}(\alpha_{[3]}-\alpha)\nonumber\\
&&\qquad\quad\;-(\beta_{[3]}-\beta)(\alpha_{[3]}-\alpha)\left[ \frac{\beta_{x}}{{k_1}+h}
+\frac{h+k_2}{2h}\beta\right]\nonumber\\
&&\qquad\quad\;+\frac{\beta_{[3]}\beta\alpha}{2h({k_1}+h)}
\left[2(\mathcal{D}\beta)(\mathcal{D}\alpha_{[3]})
-(\mathcal{D}\alpha)\mathcal{D}(\beta_{[3]}+\beta)\right]\nonumber\\
&&\qquad\quad\; +\frac{\beta_{[3]}\beta\alpha_{[3]}}{2h({k_1}+h)}
\left[2(\mathcal{D}\beta_{[3]})(\mathcal{D}\alpha)
-(\mathcal{D}\alpha_{[3]})\mathcal{D}(\beta_{[3]}+\beta)\right].
\end{eqnarray*}

While above B\"{a}cklund transformation takes a more sophisticated form than those induced by elementary Darboux transformations, it has the advantage that the reductions can be done straightforwardly as will be seen below.

\noindent
{\bf Case I.} $\beta=\alpha=\mathcal{D}\Phi$.

Let ${p_2}=-\mathrm{i}{p_1}$, $y_2=-y_1$,~$\Omega_2=\mathrm{i}\Omega_1$, then the linear problem is covariant under the DT:
\begin{eqnarray}
\label{lax21_smkdv}
\Psi_{[1]} = W\Psi,\quad
 W=
\left(
  \begin{array}{cccc}
    \lambda^2-\frac{1+y_1^2}{1-y_1^2} p_1^2 &\frac{2 p_1^2 y_1}{1-y_1^2} & \frac{2{p_1}\Omega_1}{y_1^2-1} \lambda \\[5pt]
     \frac{2 p_1^2 y_1}{y_1^2-1} & \lambda^2+\frac{1+y_1^2}{1-y_1^2} p_1^2 &  \frac{2{p_1} y_1 \Omega_1}{y_1^2-1} \lambda \\[5pt]
    \frac{2{p_1}\Omega_1}{y_1^2-1}\lambda  & \frac{2{p_1} y_1 \Omega_1}{1-y_1^2}\lambda  & \lambda^2+ p_1^2
  \end{array}
\right),
\end{eqnarray}
with
\begin{eqnarray}\label{smkdv_phi1}
 \Phi_{[1]}\equiv -\Phi+2 \mbox{arctanh}(y_1),
\end{eqnarray}
or equivalently,
\begin{eqnarray}\label{smkdv_alpha1}
\alpha_{[1]}\equiv\alpha+\frac{2{p_1}\Omega_1}{1-y_1^2}.
\end{eqnarray}
From this Darboux transformation, we may recover the Darboux transformation for the supersymmetric sine-Gordon equation constructed  in \cite{shs}.

Eliminating $y_1$ and $\Omega_1$ by (\ref{smkdv_phi1},\ref{smkdv_alpha1}), one obtains

\begin{eqnarray}\label{bd_smkdv}
\Psi_{[1]}\equiv W\Psi,
\end{eqnarray}
where
\[
 W=
\left(
  \begin{array}{cccc}
    \lambda^2-p_1^2 \cosh(\Phi+\Phi_{[1]})
    &  p_1^2 \sinh(\Phi+\Phi_{[1]})
    & \lambda{\cal D }(\Phi-\Phi_{[1]})\\
         -p_1^2 \sinh(\Phi+\Phi_{[1]})
         & \lambda^2+ p_1^2\cosh(\Phi+\Phi_{[1]})
         & -\sharp \\
    \lambda{\cal D }(\Phi-\Phi_{[1]})
    & \sharp
    &\lambda^2+ p_1^2
  \end{array}
\right)
\]
with
\[\sharp=\lambda \tanh\left(\frac{\Phi+\Phi_{[1]}}{2}\right){\cal D }(\Phi_{[1]}-\Phi).\]
Then the zero curvature equation
\begin{eqnarray*}
(\mathcal{D} W)+ W^\dag {L}-{L}_{[1]} W=0
\end{eqnarray*}
supplies the B\"{a}cklund transformation  for the supersymmetric MKdV/sinh-Gordon equation \cite{kulish,liuhu,xue}
\begin{eqnarray}\label{smkdv_bt1}
&\left(\Phi_{[1]}-\Phi\right)_x=p_1^2\sinh(\Phi+\Phi_{[1]})+\tanh \left(\frac{\Phi+\Phi_{[1]}}{2}\right)({\cal D}\Phi)({\cal D}\Phi_{[1]}).
\end{eqnarray}
 Moreover, from \eqref{smkdv_bt1} a superposition formula was obtained and it provides a convenient way  to construct multi-soliton solutions \cite{gomes,xue}.

\noindent
{\bf Case II.} $\beta=\kappa\alpha^*$.

Let~${p_2}=-\mathrm{i}{p_1^*}$, $y_2=-\kappa y_1^*$,~$\Omega_2=\mathrm{i}\kappa \Omega_1^*$, then the linear problem is covariant with respect to the following the Darboux transformation
{\small
\begin{align}
\label{laxreduction_snls}
\Psi_{[1]}&= W\Psi,\quad
 W=
\left(
  \begin{array}{cccc}
    \lambda^2+p_{11} & p_{12} & \lambda n_{13} \\
     -\kappa  p_{12}^* & \lambda^2(1+\Delta)-p_{11}^* & \lambda  n_{23} \\
    \lambda \kappa  n_{13}^* & -\lambda n_{23}^* & \lambda^2(1+\Delta)+ {p_1^*}^2 \\
  \end{array}
\right),\\
\label{field22_snls}
\alpha_{[1]}&=\alpha+\frac{2\kappa k_1 \Omega_1^*\left({p_1}+\alpha\Omega_1\right)}{|p_1|^2(1-\kappa |y_1|^2)},
\end{align}
}
where
\begin{align*}
k_1\;&= \frac{1}{2}( p_1^2+{p_1^*}^2),\qquad\;\,
\Delta\;\,=\frac{-2\kappa k_1  \Omega_1 \Omega_1^*}{|p_1|^2(1-\kappa |y_1|^2)},\\
n_{13}&=\frac{-2 \kappa k_1\Omega_1^*}{{p_1^*}(1-\kappa |y_1|^2)},\quad\;
n_{23}=y_1 n_{13} ,\\
p_{12}&= \frac{2\kappa k_1 y_1^*}{1-\kappa |y_1|^2}
       +\frac{2k_1{p_1}y_1^*\Omega_1 \Omega_1^*}{{p_1^*}(1-\kappa |y_1|^2)^2},\\
p_{11}&= -\frac{ p_1^2+\kappa{{p_1^*}}^2 |y_1|^2}{1-\kappa |y_1|^2}
       -\frac{2\kappa {p_1}k_1 \Omega_1 \Omega_1^*}{{p_1^*}(1-\kappa |y_1|^2)^2}.
\end{align*}
 We observe that if  ${p_1}$ is a pure imaginary number, then \eqref{field22_snls} coincides with the (5.19) of   \cite{rk}  constructed by Roelofs and Kersten within the framework of prolongation theory. We also remark that while they found the transformation on the level of field variables, the transformations for the eigenfunctions were missing.

Eliminating $y_1$ and $\Omega_1$, one obtains
\begin{eqnarray*}
&&n_{13}=\alpha-\alpha_{[1]}-\alpha \Delta,\\
&&n_{23}=-\frac{\kappa\mathcal{D}(\alpha_{[1]}^*-\alpha^*)}{2 h(h+{k_1})}\left[2h-\kappa\alpha(\alpha_{[1]}^*-\alpha^*)\right]\left(\alpha_{[1]}-\alpha\right),\\
&&p_{11}=-\frac{1}{2}\left(p_1^2-{{p_1^*}}^2\right)-h-\Delta_{11}-\Delta_{41},\\
&&p_{12}=\mathcal{D}(\alpha_{[1]}-\alpha)
+\frac{\kappa}{h+{k_1}}\mathcal{D}(\alpha\alpha_{[1]})(\alpha_{[1]}^*-\alpha^*)
+\frac{\kappa\Delta}{2 h}\mathcal{D}(\alpha_{[1]}-\alpha)\alpha\alpha^*,
\end{eqnarray*}
where
\begin{align*}
k_1=\;&\frac{1}{2}(p_1^2+{p_1^*}^2),\qquad h\;\equiv\sqrt{k_1^2+\kappa \mathcal{D}(\alpha_{[1]}-\alpha)\mathcal{D}(\alpha_{[1]}^*-\alpha^*)},\\
\Delta=\;&\frac{\kappa}{{k_1}+h}(\alpha_{[1]}-\alpha)(\alpha_{[1]}^*-\alpha^*),\\
\Delta_{41}=\;&\frac{h-{k_1}}{4 h^3({k_1}+h)}\left(k_1^2+2 {k_1} h+3h^2\right)\alpha\alpha^*\alpha_{[1]}\alpha_{[1]}^*,\\
\Delta_{11}=\;&
\frac{\kappa\Delta}{2h} \left[ (\mathcal{D}\alpha_{[1]}^*)(\mathcal{D}\alpha)-(\mathcal{D}\alpha_{[1]})(\mathcal{D}\alpha^*)\right]
+\frac{\kappa}{2h}\left(h-{k_1}\right)\left(\alpha_{[1]}\alpha^*+\alpha_{[1]}^*\alpha\right)
\\
&
-\frac{\kappa}{4h}\left(2h+p_1^2-{{p_1^*}}^2\right)
 \left(\alpha_{[1]}-\alpha\right)\left( \alpha_{[1]}^*-\alpha^*\right).
\end{align*}
The compatibility condition is
\begin{eqnarray*}
(\mathcal{D} W)+ W^\dag L-L_{[1]} W=0,
\end{eqnarray*}
which leads to a B\"{a}cklund transformation  for the supersymmetric NLS
\begin{align*}
\alpha_{[1],x}=&\;\alpha_x+   {p_1^*}^2\alpha_{[1]}+\alpha\left(2h+p_1^2-{p_1^*}^2\right)\left[\frac{1}{2}+\frac{\kappa}{4h}(\alpha_{[1]}^*-\alpha^*)\alpha_{[1]}\right]\nonumber\\
&+\frac{\kappa}{{k_1}+h}(\alpha_{[1]}-\alpha )\left[(\mathcal{D}\alpha_{[1]})\mathcal{D}(\alpha_{[1]}^*-\alpha^*)-(\alpha_{[1]}^*-\alpha^*)\alpha_{x}\right]\nonumber\\
&
+\frac{\alpha_{[1]}\alpha\alpha^*}{2h({k_1}+h)}\left[2(\mathcal{D}\alpha)(\mathcal{D}\alpha_{[1]}^*)
-(\mathcal{D}\alpha^*)\mathcal{D}(\alpha_{[1]}+\alpha)\right]\nonumber\\
&+\frac{\alpha_{[1]}\alpha\alpha_{[1]}^*}{2h({k_1}+h)}\left[2(\mathcal{D}\alpha_{[1]})(\mathcal{D}\alpha^*)-\mathcal{D}(\alpha_{[1]}+\alpha)(\mathcal{D}\alpha_{[1]}^*)
\right].
\end{align*}





\section{Difference systems}

In this section, we will consider the discrete systems for our supersymmetric integrable equations related to the linear problem \eqref{LAXL_snls}. Taking the Darboux and B\"{a}cklund transformations constructed  in last two sections, we will show that various discrete equations may be worked out.
To this end,  we work in the framework of components rather than the superfields and hence  we make the following expansions
\begin{eqnarray*}
\alpha=\xi+\theta q,\; \beta=\eta+\theta r, \; \Phi=v+\theta \xi, \;
\Psi=\chi+\theta b,\; \lambda_1=p_1^2,\; \lambda_2=p_2^2.
\end{eqnarray*}
It follows form  the linear problem \eqref{LAXL_snls} that
\begin{eqnarray}\label{LAXL0_snls}
\chi_x={\cal L}\chi,\quad
{\cal L}=
\left(
  \begin{array}{cccc}
    \eta \xi& q & -\lambda \xi  \\
    r & \lambda^2+\xi\eta &  0 \\
    \lambda \eta & 0  & \lambda^2
  \end{array}
\right).
\end{eqnarray}

\subsection{General SUSY AKNS }
We first study  the general case of supersymmetric AKNS system and the following three cases will be considered.

{\bf Case 1.} It follows from the eDT-I \eqref{db1_snls} that
\begin{align}\label{lax10_snls2}
\chi_{[1]}=\mathcal{W}_1\chi,\quad
\mathcal{W}_1=
\left(
  \begin{array}{cccc}
    \lambda^2+\left(1+\xi\eta_{[1]}\right) \left(q r_{[1]}-{\lambda_1}\right)
    & -q & \lambda \xi  \\
  -r_{[1]} & 1-\xi\eta_{[1]} &  0 \\
    -\lambda \eta_{[1]} & 0 & 1
  \end{array}
\right).
\end{align}
The compatibility of the linear systems (\ref{LAXL0_snls},\ref{lax10_snls2}) yields
\begin{eqnarray*}
\mathcal{W}_{1,x}+\mathcal{W}_1 \mathcal{L} -\mathcal{L}_{[1]}\mathcal{W}_1=0,
\end{eqnarray*}
which leads to a B\"{a}cklund transformation
\begin{align*}
\xi_{[1]}\;\;&=-\xi_x(1+\xi\eta_{[1]})+\xi\left(q r_{[1]}-{\lambda_1}\right),\\
\eta_{[1],x}&=\eta(1-\xi\eta_{[1]})-\eta_{[1]}\left(q r_{[1]}-{\lambda_1}\right),\\
q_{[1]}\;\;&=\left[q\left(q r_{[1]}-{\lambda_1}\right)-q_x\right](1+\xi\eta_{[1]})
+q\left(\xi_x\eta_{[1]}-\xi\eta\right),\\
r_{[1],x}&=r(1-\xi\eta_{[1]})
+r_{[1]}\left({\lambda_1}-q r_{[1]}+\xi\eta-\xi_x\eta_{[1]}\right),
\end{align*}
i.e. the component form of \eqref{snls_case1bt}. This system may be interpreted as a semi-discrete system.

{\bf Case 2.}\quad
From the eDT-II \eqref{db2_snls}, we have
\begin{align} \label{laxB1_snls}
\chi_{[2]}&={\cal W}_2\chi,\\
{\cal W}_2&=
\left(
  \begin{array}{cccc}
    1-\xi_{[2]}\eta
    & q_{[2]} & -\lambda \xi_{[2]}  \\
 r & (1+\xi_{[2]}\eta )(\lambda^2-{\lambda_2}+q_{[2]}r)
  &  -\lambda r\xi_{[2]}\\
    \lambda \eta & \lambda  q_{[2]}\eta & \lambda^2(1+\xi_{[2]}\eta )-{\lambda_2}
  \end{array}
\right).\nonumber
\end{align}
The equation
\begin{eqnarray*}
{\cal W}_{2,x}+{\cal W}_2 \mathcal{L} -\mathcal{L}_{[2]}{\cal W}_2=0
\end{eqnarray*}
yields a B\"{a}cklund transformation
\begin{align*}
\xi_{[2],x}&=\xi( {\xi_{[2]}}\eta-1 )+{\xi_{[2]}}\left(q_{[2]} r-{\lambda_2}\right),\\
\eta_{[2]}\;\;&=\eta_x(1+{\xi_{[2]}}\eta)+\eta\left(q_{[2]} r-{\lambda_2}\right),\\
q_{[2],x}&=q( {\xi_{[2]}}\eta-1 )
+q_{[2]}\left(q_{[2]} r-{\lambda_2}-\xi\eta-\xi_{[2]}\eta_x\right),\\
r_{[2]}\;\;&=\left[r_x+r\left(q_{[2]} r-{\lambda_2}\right)\right](1+{\xi_{[2]}}\eta)
-r\left(\xi\eta+\xi_{[2]}\eta_x\right),
\end{align*}
which is another semi-discrete super system. While this system may be obtained from the one given in Case 1 by a simple change of notations, we present it here since it will be useful for construction of fully discrete systems (see Case A below).


Besides these two elementary transformations eDT-I and eDT-II, we may work {\it directly} with the spectral problem \eqref{LAXL0_snls} and consider its Darboux transformation. In this way, we find the following.

{\bf Case 3.}
\begin{eqnarray}
\label{laxC20_snls}
\chi_{[2]}\equiv \mathcal{N}\chi,\quad
\mathcal{N}=
\left(
  \begin{array}{cccc}
    \lambda^2+F
    & -q & \lambda \xi  \\
  -r_{[2]} & 0 &  0 \\
    -\lambda \eta_{[2]} & 0 & c
  \end{array}
\right),
\end{eqnarray}
where \begin{eqnarray}\label{F}
F=(\ln q)_x+\xi_{[2]}\eta_{[2]}+\xi\eta,\quad r_{[2]}=l/{q},
\end{eqnarray} $c$ and $l$ are even constants.
The  condition
\begin{eqnarray*}
\mathcal{N}_x+\mathcal{N} \mathcal{L} -\mathcal{L}_{[2]}\mathcal{N}=0
\end{eqnarray*}
is equivalent to
\begin{subequations}\label{toda}
\begin{align}
 \xi_x\;\;\;\;\;&=(\ln q)_x\xi-c\xi_{[2]},\\
 \eta_{[2],x}\;\;\;&=-(\ln q)_x\eta_{[2]}+c \eta,\\
(\ln q)_{xx}&=q r-q_{[2]}r_{[2]} -\xi_{[2],x}\eta_{[2]}-\xi\eta_x.
\end{align}
\end{subequations}
Introducing new variable $v$ via $q=e^{-v}$, one has
\begin{align*}
 \xi_x\;\;\;\,&=-v_x\xi-c\xi_{[2]},\nonumber\\
 \eta_{[2],x}&= v_x\eta_{[2]}+c \eta,\nonumber\\
v_{xx}\;\;&=l(e^{v-v_{[2]}}-e^{v_{[-2]}-v}) +\xi_{[2],x}\eta_{[2]}+\xi\eta_x,
\end{align*}
which is the fermionic extension of Toda lattice.
In the case of $c=0$, it follows from (\ref{toda}a,\ref{toda}b) that \[(\xi /q)_x=0,\quad (\eta_{[2]}q)_x=0,\]
which leads to
\[
v_{xx}=l(e^{v-v_{[2]}}-e^{v_{[-2]}-v})+\mu_1\mu_2\left(v_{[-2],x}e^{v_{[-2]}-v}-v_{[2],x}e^{v-v_{[2]}}\right),
\]
where $\mu_k (k=1, 2)$ are odd constants.


Above we reinterpreted the B\"acklund transformations resulted from the  Darboux transformations and obtained three semi-discrete super systems. Next we aim to construct some  fully discrete systems.

{\bf Case A.}
Consider two Darboux transformations (\ref{lax10_snls2},\ref{laxB1_snls}). Their compatibility yields~
\begin{subequations}\label{dde2_snls}
\begin{align}
\xi_{[12]} &=\xi-\frac{({\lambda_1}-{\lambda_2})}{(1+q_{[2]}r_{[1]})^2}
       \xi_{[2]}\left(1+q_{[2]}r_{[1]}+\xi\eta_{[1]}\right),\\
\eta_{[21]}&=\eta+\frac{({\lambda_1}-{\lambda_2})}{(1+q_{[2]}r_{[1]})^2}
       \eta_{[1]}\left(1+q_{[2]}r_{[1]}+\xi_{[2]}\eta\right),\\
q_{[12]}&=q+\frac{{\lambda_2}-{\lambda_1}}{1+q_{[2]}r_{[1]}}\left(q_{[2]}-q\xi_{[2]}\eta_{[1]}\right)-\frac{({\lambda_1}-{\lambda_2})q_{[2]}}{(1+q_{[2]}r_{[1]})^2}\nonumber\\
  &\quad\times\left( \xi_{[2]}\eta +(\xi+{\lambda_1}\xi_{[2]})\eta_{[1]}
  +\frac{1-q_{[2]}r_{[1]}}{1+q_{[2]}r_{[1]}}\xi_{[2]}\xi\eta_{[1]}\eta\right),\\
  r_{[21]}&=r+\frac{{\lambda_1}-{\lambda_2}}{1+q_{[2]}r_{[1]}}\left(r_{[1]}-r\xi_{[2]}\eta_{[1]}\right)+\frac{({\lambda_1}-{\lambda_2})r_{[1]}}{(1+q_{[2]}r_{[1]})^2}\nonumber\\
  &\quad\times\left( \xi_{[2]}\eta +(\xi+{\lambda_2}\xi_{[2]})\eta_{[1]}
  +\frac{1-q_{[2]}r_{[1]}}{1+q_{[2]}r_{[1]}}\xi_{[2]}\xi\eta_{[1]}\eta\right).
\end{align}
\end{subequations}




{\bf Case B.}
By considering the compatibility of (\ref{lax10_snls2},\ref{laxC20_snls}), or
\begin{eqnarray*}
\mathcal{W}_{1,[2]}\mathcal{N}=\mathcal{N}_{[1]}\mathcal{W}_1,
\end{eqnarray*}
leads to the following system
\begin{align}
\xi_{[1]}&=c\xi_{[2]}+\frac{q_{[1]}}{q}\xi,\nonumber\\
\eta_{[2]}&=c\eta_{[1]}+\frac{q_{[1]}}{q}\eta_{[12]},\nonumber\\
F\;&=(q r_{[1]}-{\lambda_1})(1+\xi\eta_{[1]})-\frac{q_{[1]}}{q}\left(1-\xi_{[2]}\eta_{[12]}\right),\label{F_snls}\\
F_{[1]}&=\left({q_{[2]}}{r_{[12]}}-{\lambda_1}\right)(1+\xi_{[2]}\eta_{[12]})-\frac{q_{[1]}}{q}\left(1-\xi\eta_{[1]}\right),\label{F2_snls}
\end{align}
where $F$ is given by \eqref{F} and $r_{[1]}={l}/{q_{[1,-2]}}$.
By shifting $F_{[1]}$, it follows from (\ref{F_snls},\ref{F2_snls}) that
\begin{eqnarray*}
\lefteqn{\left( \frac{lq}{q_{[1,-2]}}-{\lambda_1}\right)\left(1+\xi\eta_{[1]}\right)-\frac{q_{[1]}}{q}\left(1-\xi_{[2]}\eta_{[12]}\right)=}\nonumber\\
\;\;\;&&\qquad\left(\frac{ l q_{[-1,2]}}{q}-\lambda_{1}\right)\left(1+\xi_{[-1,2]}\eta_{[2]}\right)-\frac{q}{q_{[-1]}}\left(1-\xi_{[-1]}\eta\right).
\end{eqnarray*}
Let $q=e^{-v}$, we have
\begin{align}
&\xi_{[1]}=c\xi_{[2]}+e^{v-v_{[1]}}\xi,\nonumber\\
&\eta_{[2]}=c\eta_{[1]}+e^{v-v_{[1]}}\eta_{[12]},\nonumber\\
&{v_x}\;\,=
({\lambda_1}- l e^{v_{[1,-2]}-v})(1+\xi\eta_{[1]})+e^{v-v_{[1]}}+\xi\eta+c\xi_{[2]}\eta_{[1]},\nonumber\\
&\left(l e^{v_{[1,-2]}-v}-{\lambda_1}\right)\left(1+\xi\eta_{[1]}\right)-e^{v-v_{[1]}}\left(1-\xi_{[2]}\eta_{[12]}\right)=\nonumber\\
&\quad\;\,\left(le^{v-v_{[-1,2]}}-\lambda_{1}\right)\left(1+\xi_{[-1,2]}\eta_{[2]}\right)-e^{v_{[-1]}-v}\left(1-\xi_{[-1]}\eta\right).\label{snls_di1toda}
\end{align}
In the bosonic limit,  \eqref{snls_di1toda} reduces to the discrete Toda lattice
\cite{toda1975} or Hirota equation \cite{hirota19772}
\begin{eqnarray*}
l (e^{v_{[1,-2]}-v}-e^{v-v_{[-1,2]}})=e^{v-v_{[1]}}-e^{v_{[-1]}-v}.
\end{eqnarray*}
Thus, \eqref{snls_di1toda} may be regarded as a super discrete Toda lattice.


\subsection{SUSY MKdV}


Now we turn to the supersymmetric sinh-Gordon/MKdV equations and manage to find their discretizations.  In this case, $\xi=\eta$ and $q=r=v_x$. From \eqref{bd_smkdv}, one obtains
{\small\begin{align}\label{smkdv_lax1}
    & \chi_{[1]}=\mathcal{W} \chi ,\\
    & \mathcal{W} \;\,=\nonumber\\
&
\left(
  \begin{array}{cccc}
    \lambda^2-\lambda_1 \cosh(v +v_{[1]})  &  \lambda_1 \sinh(v +v_{[1]})
    & \lambda(\xi-\xi_{[1]})\\
         -\lambda_1 \sinh(v +v_{[1]})     & \lambda^2+\lambda_1\cosh(v +v_{[1]})
         & \lambda \tanh\left(\frac{v +v_{[1]}}{2}\right)(\xi-\xi_{[1]}) \\
    \lambda(\xi-\xi_{[1]})  & \lambda \tanh\left(\frac{v +v_{[1]}}{2}\right)(\xi_{[1]}-\xi)   &\lambda^2+ \lambda_1
  \end{array}
\right).\nonumber
\end{align}
}
Now the compatibility condition of  (\ref{LAXL0_snls}) and \eqref{smkdv_lax1}, namely
\begin{eqnarray*}
\mathcal{W}_x+\mathcal{W} \mathcal{L} -\mathcal{L}_{[1]}\mathcal{W}=0
\end{eqnarray*}
holds if and only if 
\begin{subequations}\label{smkdv_btf}
\begin{align} 
 (\xi_{[1]}-\xi )_x&={\lambda_1}\big[\xi +\xi_{[1]}\cosh(v +v_{[1]})\big]+v_{x}\tanh\left(\frac{v +v_{[1]}}{2}\right) (\xi_{[1]}-\xi ),\\
 (v_{[1]}-v )_x&={\lambda_1} \sinh(v +v_{[1]})+\xi \xi_{[1]} \tanh\left(\frac{v +v_{[1]}}{2}\right) 
\end{align}
\end{subequations}
are satisfied.
Above equations are the components-version of \eqref{smkdv_bt1}.
Then, we consider one transformation \eqref{smkdv_lax1} and the other
\begin{eqnarray}\label{smkdv_lax2}
\chi_{[2]}={\cal V}\chi,
\end{eqnarray}
where the matrix ${\cal V}$ is the matrix ${\cal W}$ of \eqref{smkdv_lax1}
with $\lambda_1$,  $\xi_{[1]}$ and $v_{[1]}$ replaced by $\lambda_2$,  $\xi_{[2]}$ and $v_{[2]}$ respectively.
Let $ v=\ln(w)$, the compatibility condition of \eqref{smkdv_lax1} and \eqref{smkdv_lax2}, namely
\begin{eqnarray*}
{\cal W}_{[2]}{\cal V}={\cal V}_{[1]}{\cal W}
\end{eqnarray*}
yields an integrable difference system
\begin{subequations}\label{dde_smkdv}
\begin{align}\label{smkdv_xi12}
\xi_{[12]}=&\;
\xi
+\frac{{\lambda_1}+{\lambda_2}}{1+w w_{[1]}}
 \left(\frac{w_{[1]}}{{\lambda_1} w_{[1]}-{\lambda_2} w_{[2]}}
       +\frac{ww_{[1]} w_{[2]}}{{\lambda_1} w_{[2]}-{\lambda_2} w_{[1]}}
 \right)
 \left(\xi_{[1]}-\xi\right) \nonumber\\
&+\frac{{\lambda_1}+{\lambda_2}}{1+w w_{[2]}}
 \left(\frac{w_{[2]}}{{\lambda_2} w_{[2]}-{\lambda_1} w_{[1]}}
       +\frac{w w_{[1]}w_{[2]}}{{\lambda_2} w_{[1]}-{\lambda_1} w_{[2]}}
 \right)
 \left(\xi_{[2]}-\xi\right),\\
w_{[12]}=
&\;\frac{{\lambda_1}w_{[1]}-{\lambda_2}w_{[2]}}{{\lambda_1}w_{[2]}-{\lambda_2}w_{[1]}} w -\left(\xi_{[1]}-\xi\right)\left(\xi_{[2]}-\xi\right)\nonumber\\
&\times\frac{2({\lambda_1}+{\lambda_2})w^2w_{[1]} w_{[2]} (w_{[1]}-w_{[2]})}{(1+w w_{[1]})(1+w w_{[2]})({\lambda_1}w_{[2]}-{\lambda_2}w_{[1]})^2}
   .\label{smkdv_u12}
\end{align}
\end{subequations}


It is noticed that with the absence of fermionic variables, \eqref{dde_smkdv} reduces to the well-known lattice potential MKdV equation \cite{Nijihoff}
\begin{eqnarray*}
w_{[12]}=\frac{{\lambda_1}w_{[1]}-{\lambda_2}w_{[2]}}{{\lambda_1}w_{[2]}-{\lambda_2}w_{[1]}}w,
\end{eqnarray*}
thus \eqref{dde_smkdv} constitutes a super extension of the lattice potential MKdV system.

\section{Continuum limits}
In last section, we constructed some semi-discrete and fully discrete systems. To gain a better understanding of a discrete system, one may consider its continuum limits (see \cite{Nijihoff} for example). The aim of this section is to work out the continuum limits of the discrete systems \eqref{smkdv_btf}  and \eqref{dde_smkdv} above. To this end and for convenience, by employing the symmetries of supersymmetric MKdV equation, we keep $w,\;\xi,\;w_{[12]}$ and $\xi_{[12]}$ unchanged£¬ but replace  $w_{[i]}, v_{[i]}$ and $\xi_{[i]}$  by $w_{[i]}^{-1}$, $-v_{[i]}$ and $-\xi_{[i]}$ ($i=1,\;2$) respectively. In this way, from
\eqref{smkdv_btf}  and \eqref{dde_smkdv} we have the  differential-difference system
\begin{subequations}\label{bt0_smkdv}
\begin{align} 
 (\xi_{[1]}+\xi )_x&={\lambda_1}\big[\xi_{[1]}\cosh(v_{[1]}-v )-\xi \big]-v_{x}(\xi_{[1]}+\xi )\tanh\left(\frac{v_{[1]}-v }{2}\right) ,\\
 (v_{[1]}+v )_x&={\lambda_1} \sinh(v_{[1]}-v)-\xi \xi_{[1]} \tanh\left(\frac{v_{[1]}-v }{2}\right), 
\end{align}
\end{subequations}
and the difference-difference system
\begin{subequations}\label{dde0_smkdv}
\begin{align}\label{smkdv_xi120}
0=\,&(w_{[1]}+w )(w_{[2]}+w )({\lambda_1} w_{[2]}-{\lambda_2} w_{[1]})({\lambda_1} w_{[1]}-{\lambda_2} w_{[2]})(\xi-\xi_{[12]})\nonumber\\
&-({\lambda_1}+{\lambda_2})(w_{[2]}+w )w_{[1]}\left(\xi_{[1]}+\xi\right)\nonumber\\
&\times\left[ w_{[2]}({\lambda_1} w_{[1]} -{\lambda_2} w_{[2]})+w ({\lambda_1} w_{[2]}-{\lambda_2} w_{[1]})\right]\nonumber\\
&  +({\lambda_1}+{\lambda_2})(w_{[1]}+w ) w_{[2]} \left(\xi_{[2]}+\xi\right)\nonumber\\
&\times\left[ w_{[1]}({\lambda_1} w_{[1]}-{\lambda_2} w_{[2]})+w ({\lambda_1} w_{[2]}-{\lambda_2} w_{[1]})\right]
,\\
0=\,&
(w_{[1]}+w )(w_{[2]}+w )({\lambda_1} w_{[1]}-{\lambda_2} w_{[2]})\nonumber\\
&\times
\left[w({\lambda_1} w_{[2]}-{\lambda_2} w_{[1]}) -w_{[12]}({\lambda_1} w_{[1]}-{\lambda_2}w_{[2]})
\right]\nonumber\\
&-2({\lambda_1}+{\lambda_2})w^2 w_{[1]}w_{[2]}(w_{[2]}-w_{[1]})
   \left(\xi_{[1]}+\xi\right)\left(\xi_{[2]}+\xi\right).\label{smkdv_u120}
\end{align}
\end{subequations}
These systems are the discrete versions of the potential supersymmetric MKdV equation.
 We now justify this claim by considering their continuum limits.
\subsection{The continuum limit of \eqref{bt0_smkdv} }\label{xiuhatp}
To effect the continuous limit of \eqref{bt0_smkdv} we introduce a new continuous variable $\tau$ by  $\tau=\frac{2n_1}{\lambda_1} $, and
\[
\xi\equiv \xi(x,\tau),\quad v\equiv v(x,\tau),\quad\xi_{[1]}\equiv \xi\left(x,\tau+\frac{2}{\lambda_1}\right),\;\;v_{[1]}\equiv v\left(x,\tau+\frac{2}{\lambda_1}\right). 
\]
Now expanding the quantities in \eqref{bt0_smkdv} in powers of $\frac{1}{\lambda_1}$, and defining a new independent variable  $t$ in term of $\tau$ and $x$ such that
\begin{eqnarray*}
\partial_\tau=\partial_x+\frac{1}{3\lambda_1^2}\partial_t,
\end{eqnarray*}
 we obtain in the continuous limit up to terms of order $\frac{1}{\lambda_1}$
\begin{eqnarray}\label{psmkdv}
\xi_t=\xi_{xxx}-3v_x^2\xi_x-3v_xv_{xx}\xi,\quad
v_t=v_{xxx}-2v_x^3+3v_x\xi\xi_x,
\end{eqnarray}
which is the potential supersymmetric MKdV equation in component form.

\subsection{The semi-continuous limits of  \eqref{dde0_smkdv}}
This system is a fully discrete system and different continuous limits are meaningful. We consider in this subsection the semi-continuous limits which consist of the straight continuous limit and skew continuous limit. In the first case, we obtain the differential-difference system \eqref{bt0_smkdv} as it should be, and in the second case a new semi-discrete system is derived.

\subsubsection{Straight continuum limit}
To see that the system \eqref{dde0_smkdv}  may be regarded as a discretization
of the differential-difference system \eqref{bt0_smkdv}, we define
\begin{align*}
\xi\;\;&\equiv  \xi_{n_1}(x),\quad w\equiv w_{n_1}(x),\quad x=\frac{2n_2}{\lambda_2},\\
\xi_{[2]} &\equiv \xi_{n_1}\left(x+\frac{2}{\lambda_2}\right),\quad\;\;
\xi_{[12]}\equiv\xi_{n_1+1}\left(x+\frac{2}{\lambda_2}\right),\\
w_{[2]}&\equiv w_{n_1}\left(x+\frac{2}{\lambda_2}\right),\quad
w_{[12]}\equiv w_{n_1+1}\left(x+\frac{2}{\lambda_2}\right).
\end{align*}
Applying the Taylor expansions
\begin{eqnarray*}
&\xi_{n}\left(x+\frac{2}{\lambda_2}\right)=\xi_{n}+\frac{2}{\lambda_2}\xi_{n,x}+O\left(\frac{1}{\lambda_2^2}\right),\\
&w_{n}\left(x+\frac{2}{{\lambda_2}}\right)=w_{n}+\frac{2}{{\lambda_2}}w_{n,x}+O\left(\frac{1}{\lambda_2^2}\right),
\end{eqnarray*}
where $n$ can take the value $n_1$ or $n_1+1$, and substituting them into \eqref{dde0_smkdv} and letting $w_{n_1}=e^{v_{n_1}}$ and $w_{n_1+1}=e^{v_{n_1+1}}$, the leading terms yield
\begin{subequations}\label{bt_smkdv}
\begin{align} 
(\xi_{n_1+1}+\xi_{n_1} )_x&={\lambda_1}\big[\xi_{n_1+1}\cosh(v_{n_1+1}-v_{n_1} )-\xi_{n_1} \big] \nonumber\\
&\quad -v_{n_1,x}(\xi_{n_1+1}+\xi_{n_1} )\tanh\left(\frac{v_{n_1+1}-v_{n_1} }{2}\right) ,\\
 (v_{n_1+1}+v_{n_1} )_x&= {\lambda_1} \sinh( v_{n_1+1}-v_{n_1})\nonumber\\
 &\quad-\xi_{n_1} \xi_{n_1+1} \tanh\left(\frac{v_{n_1+1}-v_{n_1} }{2}\right) ,
\end{align}
\end{subequations}
i.e. \eqref{bt0_smkdv}.

\subsubsection{Skew continuum limit}

To perform this continuum limit, we introduce
\begin{eqnarray} \label{skew}
N=n_1+n_2+1,\quad \lambda_2=\lambda_1+\epsilon ,\quad \tau=\epsilon n_2,
\end{eqnarray}
and
\begin{align} \label{skew1}
\xi\;\;\; &\equiv \xi_{N-1}(\tau),\qquad \quad\; w\;\;\;\equiv w_{N-1}(\tau),\\ \nonumber
\xi_{[1]}\; &\equiv \xi_N,\qquad \qquad\quad\; w_{[1]}\;\equiv w_{N}(\tau),\\ \nonumber
\xi_{[2]}\; &\equiv \xi_{N}(\tau+\epsilon),\quad\quad\; w_{[2]}\;\equiv w_{N}(\tau+\epsilon),\\ \nonumber
\xi_{[12]} &\equiv \xi_{N+1}(\tau+\epsilon),\quad \,w_{[12]}\equiv w_{N+1}(\tau+\epsilon).
\end{align}
Then taking account of the Taylor expansions in $\epsilon $, the leading order terms of  \eqref{dde0_smkdv} give
\begin{subequations}\label{newdde}
\begin{align}
0=&-2\lambda_1w_N[\lambda_1w_{N,\tau}(w_{N}^2-w_{N-1}^2)+w_{N}(w_{N}+w_{N-1})^2]\xi_{N,\tau}\nonumber \\ \nonumber
&+[\lambda_1w_{N,\tau}(w_{N}-w_{N-1})+w_{N}(w_{N}+w_{N-1})]^2(\xi_{N}+\xi_{N-1})\\
&+(w_{N}+w_{N-1})^2(\lambda_1^2w_{N,\tau}^2-w_{N}^2)(\xi_{N}+\xi_{N+1}),\\ \nonumber
0=&\;4\lambda_1w_{N}^2w_{N-1}^2(\xi_{N}+\xi_{N-1})\xi_{N,\tau} w_{N,\tau}\\ \nonumber
&+(w_{N}+w_{N-1})^2(\lambda_1w_{N,\tau}+w_{N} )\nonumber\\
&\times[\lambda_1w_{N,\tau}(w_{N+1}+w_{N-1})+w_{N}(w_{N+1}-w_{N-1})],
\end{align}
\end{subequations}
which in terms of $\xi_{N,\tau}, w_{N,\tau}$ may be reformulated as
{\small
\begin{subequations}\label{newdde2}
\begin{align}
\xi_{N,\tau}&=\frac{w_{N-1}(w_{N}+w_{N+1})(\xi_{N}+\xi_{N-1})}{\lambda_1(w_{N-1}+w_{N+1})(w_{N}+w_{N-1})}
-\frac{w_{N+1}(w_{N}+w_{N-1})(\xi_{N}+\xi_{N+1})}{\lambda_1(w_{N-1}+w_{N+1})(w_{N}+w_{N+1})}, \label{tau1a}\\
 w_{N,\tau}&=\frac{w_N(w_{N-1}-w_{N+1})}{\lambda_1(w_{N-1}+w_{N+1})}\nonumber\\
 &\quad\times
 \left[1+\frac{2w_Nw_{N-1}w_{N+1}(\xi_{N}+\xi_{N-1})(\xi_{N}+\xi_{N+1})}{\lambda_1(w_{N-1}+w_{N+1})(w_{N}+w_{N+1})(w_{N-1}+w_{N})}\right]
, \label{tau1b}
\end{align}
\end{subequations}
}
or the trivial solution
\begin{eqnarray*}
\xi_{N,\tau}=\frac{w_{N-1}(\xi_{N}+\xi_{N-1})}{\lambda_1(w_{N}+w_{N-1})},\quad w_{N,\tau}=-\frac{w_{N}}{\lambda_1}.
\end{eqnarray*}

If the fermionic variable $\xi$ is absent, the system \eqref{newdde2} reduces to
\begin{align*}
\lambda_1 \partial_\tau \ln w_{N}&=\frac{w_{N-1}-w_{N+1}}{w_{N-1}+w_{N+1}},
\end{align*}
which is (4.2.19) of \cite{Nijihoff}.

\subsection{Full continuum limit}
Above the semi-continuous limits were analyzed for the fully discrete system  \eqref{dde0_smkdv} and  two differential-difference systems, namely \eqref{bt_smkdv} and \eqref{newdde2}, were obtained. The first one is \eqref{bt0_smkdv} and Subsection \ref{xiuhatp} demonstrated that its continuous limit leads to the potential supersymmetric MKdV equation.  In the remaining part of this section, we will show that  the continuum limit of \eqref{newdde2} also leads to the potential  supersymmetric MKdV equation.

Indeed, defining
\begin{eqnarray*}
\xi_{N}(\tau)\equiv \xi(s,\tau),\quad w_{N}(\tau)\equiv w(s,\tau),\quad s=\frac{2N}{\lambda_1},
\end{eqnarray*}
and
\begin{eqnarray*}
\xi_{N\pm1}(\tau)\equiv \xi\left(s\pm\frac{2}{\lambda_1},\tau\right),\quad w_{N\pm1}(\tau)\equiv w\left(s\pm\frac{2}{\lambda_1},\tau\right),
\end{eqnarray*}
then expanding \eqref{newdde} in $\frac{1}{\lambda_1}$, after cancellations the leading terms provide
\begin{subequations}\label{final_limit}\begin{align}
\xi_t&=\xi_{xxx}-\frac{3w_x^2}{w^2}\xi_x+3\left(\frac{w_x^3}{w^3}-\frac{w_xw_{xx}}{w^2}\right)\xi,\\
w_t&=w_{xxx}-\frac{3w_x}{w}w_{xx}+3w_x\xi\xi_x,
\end{align}\end{subequations}
where the new independent variables $x,\, t$ are defined by
\begin{eqnarray*}
&\partial_s=\partial_x,\quad\partial_\tau=-\frac{2}{\lambda_1^2}\partial_x-\frac{4}{3\lambda_1^4}\partial_t.
\end{eqnarray*}

An easy calculation shows that the system \eqref{final_limit} may be brought to the potential form of supersymmetric MKdV equation \eqref{psmkdv} by $w=e^{v}$.

\section{Conclusion}
Darboux transformations for a supersymmetric AKNS problem have been  constructed. They may serve a useful tool to study the associated integrable systems. In the present paper, we employed these Darboux transformations to establish discretizations for supersymmetric integrable systems and both semi-discrete and fully discrete systems are obtained. In particular, a super discrete Toda lattice and a super lattice potential MKdV systems are found. Different continuum limits are studied for the super lattice potential MKdV systems. We could go further to construct the discrete versions for the supersymmetric NLS system, whose explicit form would be rather intricate,  so we do not put them here.

Compared with the classical integrable systems or soliton equations, the study of both Darboux transformations and discretizations is still in its infancy and there is much more to be understood. For instance, it would be interesting to construct Darboux transformations for the fully supersymmetric AKNS problem of Morosi and Pizzocchero \cite{mp}. Finding proper ways to integrable discretizations for  $N=2$ supersymmetric integrable systems are also worth investigating.


\section*{Acknowledgments}

The constructive comments and suggestions of the anonymous referees have been useful for improving this paper. This work is supported by the National Natural Science Foundation of China (grant numbers:  11271366, 11331008 and 11401572) and the Fundamental Research Funds for Central Universities.


\end{document}